\documentclass[aps,preprint,showpacs,showkeys,nofootinbib,floatfix]{revtex4}
\usepackage{epsfig}
\usepackage{graphicx}
\begin{document} 

\title{\textbf{$J=3/2$ charmed hypertriton}}
\author{H.~Garcilazo} 
\email{humberto@esfm.ipn.mx} 
\affiliation{Escuela Superior de F\' \i sica y Matem\'aticas, \\ 
Instituto Polit\'ecnico Nacional, Edificio 9, 
07738 M\'exico D.F., Mexico} 

\author{A.~Valcarce} 
\email{valcarce@usal.es} 
\affiliation{Departamento de F\'\i sica Fundamental and IUFFyM,\\ 
Universidad de Salamanca, E-37008 Salamanca, Spain}

\author{T.F.~Caram\'es} 
\email{valcarce@usal.es} 
\affiliation{Departamento de F\'\i sica Fundamental and IUFFyM,\\ 
Universidad de Salamanca, E-37008 Salamanca, Spain}

\date{\today} 

\begin{abstract}
By solving exact three-body equations, we study the three-baryon system 
with charm $+1$. We look for possible bound states using baryon-baryon interactions
obtained from a chiral constituent quark model.
The smaller effect of the $\Lambda_c \leftrightarrow \Sigma_c$ conversion
reverses the order of the $(I,J)=(0,1/2)$ and $(I,J)=(0,3/2)$ states, rather
close on the strange sector. The diminishing of the kinetic energy due to the
large reduced mass gives rise to a bound state in the $(I,J)=(0,3/2)$
channel. After correcting for Coulomb effects the binding energy would be 
between 140 and 715 keV.

\end{abstract}

\pacs{21.45.-v,25.10.+s,12.39.Jh}
\keywords{baryon-baryon interactions, Faddeev equations} 
\maketitle 

\begin{figure*}[b]
\vspace*{-1cm}
\resizebox{12.cm}{15.cm}{\includegraphics{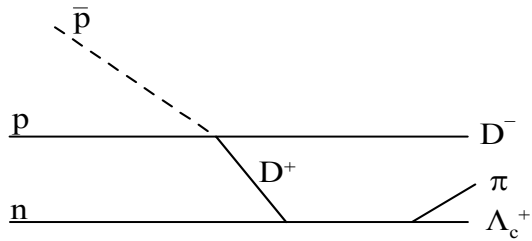}}
\vspace*{-10.5cm}
\caption{Diagram to generate $\Lambda_c^+$ by means of antiproton collisions
on the deuteron.}
\label{fig0}
\end{figure*}

Soon after the discovery of baryons possessing net charm 
it was suggested that there should also exist charmed nuclei.
The observation of a candidate event that could be interpreted
in terms of the decay of a charmed nucleus~\cite{Tip75}, fostered conjectures
about the  possible existence of charm analogs of strange 
hypernuclei~\cite{Dov77,Iwa77,Gat78}. Three ambiguous candidates of charmed hypernuclei were reported 
by an emulsion experiment with 250 GeV protons~\cite{Bat81}. 
This gave rise to several theoretical estimates about the binding
energies and the potential-well depth of charmed 
hypernuclei based on one-boson-exchange potentials for the
charmed baryon-nucleon potential~\cite{Bha81,Ban82,Ban83,Gib83,Sta86}.
There were also theoretical estimations of the production cross-sections
as well as experimental condition requirements for producing charmed
hypernuclei by means of charmed exchange reactions on nuclei~\cite{Bre89}.
The experiments for searching charmed hypernuclei are becoming realistic
and may be performed in coming years at Hall C of JPARC~\cite{Tam12},
at the FAIR experiment~\cite{Wie11}, or at the Super$B$ collider~\cite{Fel12}.
All these experimental prospects have reinvigorated twenty years later the
study of charmed hypernuclei~\cite{Cai03,Tsu04,Kop07} and also more recently theoretical
studies of charmed dibaryons~\cite{Liu12,Hua13}. We show in Fig.~\ref{fig0} a diagram to
generate $\Lambda_c^+$ baryons by means of antiproton collisions on the deuteron 
through the intermediate production of charged $D$ mesons ($m_{D^\pm}=$ 1869.61 MeV/c$^2$) feasible
at modern factories, that has been proposed to study the existence of 
charmed hypernuclei at JPARC~\cite{Tam12}. A similar reaction, $D^+ \, + \, p \, \to \,
\Lambda_c^+ \, + \pi^+$, is proposed at Super$B$ to 
detect charmed super-nuclei~\cite{Fel12}.

We have developed in the past the exact formalism to study 
three-baryon systems with a heavy flavor baryon~\cite{Gar07}.
It is our purpose in this work to extend our previous study to
three-baryon systems with a unit of charm by considering the region 
of the $\Lambda_c NN$ bound states, where a charmed hypertriton 
might exist, and calculating for the first time the $\Lambda_c d$ 
and $\Sigma_c d$ scattering lengths. We will 
simultaneously study all $\Lambda_c NN$ and $\Sigma_c NN$ states with $J=1/2,3/2$
and $I=0,1,2$. 

Let us start by summarizing the description of the three-baryon system
with a charmed baryon.
We transform the Faddeev equations from being integral 
equations in two continuous variables into integral equations in just one
continuous variable by expanding the two-body $t-$matrices
in terms of Legendre polynomials, $P_m$~\cite{Ter06},
\begin{equation}
t_{i;s_ii_i}(p_i,p^\prime_i;e)=\sum_{nr}P_n(x_i)
\tau_{i;s_ii_i}^{nr}(e)P_r(x^\prime_i),
\label{for11}
\end{equation}
where $x_{i}={\frac{p_{i}-b}{p_{i}+b}}$, 
$x_{i}^\prime={\frac{p_{i}^\prime-b}{p_{i}^\prime+b}}$, 
$p_i$ and $p_i^\prime$ are the initial and final relative momenta of
the pair $jk$, and $b$ is a scale parameter of which the results do not
depend on.

If we identify particle 1 with the charmed baryon and particles 2 and 3 
with the two nucleons, the integral
equations for $\beta d$ scattering at threshold, with $\beta=\Sigma_c$
or $\Lambda_c$, are in the case of pure $S-$wave configurations,
\begin{eqnarray}
T_{2;SI;\beta}^{ns_2i_2}(q_2) & = & 
B_{2;SI;\beta}^{ns_2i_2}(q_2)  
+\sum_{ms_3i_3} \int_0^\infty dq_3\,\left[
(-1)^{1+\sigma _{1}+\sigma _{3}-s_{2}+\tau _{1}+\tau _{3}-i_{2}} 
A_{23;SI}^{ns_2i_2ms_3i_3}(q_2,q_3;E) \right.
\nonumber \\
& + &  \left. 2\sum_{rs_1i_1} \int_0^\infty dq_1\,
A_{31;SI}^{ns_2i_2rs_1i_1}(q_2,q_1;E)
A_{13;SI}^{rs_1i_1ms_3i_3}(q_1,q_3;E) \right]
T_{2;SI}^{ms_3i_3}(q_3),
\label{for20}
\end{eqnarray}
where $\sigma _{1}$ ($\tau _{1})$and $\sigma _{3}$ ($\tau _{3})$ stand for
the spin (isospin) of the charmed baryon and the nucleon respectively, while
$s_i$ and $i_i$ are the spin and isospin of the pair $jk$.
$T_{2;SI;\beta}^{ns_2i_2}(q_2)$ is a two-component vector,
\begin{equation}
T_{2;SI;\beta}^{ns_2i_2}(q_2) = \left( \matrix{
T_{2;SI;\Sigma_c\beta}^{ns_2i_2}(q_2)  \cr
T_{2;SI;\Lambda_c\beta}^{ns_2i_2}(q_2) \cr } \right),
\label{for2}
\end{equation}
while the kernel of Eq.~(\ref{for20}) is a $2\times 2$ matrix defined by
\begin{eqnarray}
A_{23;SI}^{ns_2i_2ms_3i_3}(q_2,q_3;E)&=&\left(\matrix{
A_{23;SI;\Sigma_c\Sigma_c}^{ns_2i_2ms_3i_3}(q_2,q_3;E)&
A_{23;SI;\Sigma_c\Lambda_c}^{ns_2i_2ms_3i_3}(q_2,q_3;E)\cr
A_{23;SI;\Lambda_c\Sigma_c}^{ns_2i_2ms_3i_3}(q_2,q_3;E)&
A_{23;SI;\Lambda_c\Lambda_c}^{ns_2i_2ms_3i_3}(q_2,q_3;E)\cr }\right), \nonumber \\
A_{31;SI}^{ns_2i_2rs_1i_1}(q_2,q_1;E)&=&\left(\matrix{
A_{31;SI;\Sigma_c N(\Sigma_c)}^{ns_2i_2rs_1i_1}(q_2,q_1;E)&
A_{31;SI;\Sigma_c N(\Lambda_c)}^{ns_2i_2rs_1i_1}(q_2,q_1;E)\cr
A_{31;SI;\Lambda_c N(\Sigma_c)}^{ns_2i_2rs_1i_1}(q_2,q_1;E)&
A_{31;SI;\Lambda_c N(\Lambda_c)}^{ns_2i_2rs_1i_1}(q_2,q_1;E)\cr}\right), \nonumber \\
A_{13;SI}^{rs_1i_1ms_3i_3}(q_1,q_3;E)  &=&\left(\matrix{
A_{13;SI;N\Sigma_c}^{rs_1i_1ms_3i_3}(q_1,q_3;E)  &
   0 \cr
   0&
A_{13;SI;N\Lambda_c}^{rs_1i_1ms_3i_3}(q_1,q_3;E)  \cr}\right) \, ,
\label{gl4}
\end{eqnarray}
where
\begin{eqnarray}
A_{23;SI;\alpha\beta}^{ns_{2}i_{2}ms_{3}i_{3}}(q_{2},q_{3};E)
&=&h_{23;SI}^{s_{2}i_{2}s_{3}i_{3}}\sum_{r}\tau
_{2;s_{2}i_{2};\alpha\beta}^{nr}(E-q_{2}^{2}/2\nu _{2}){\frac{q_{3}^{2}}{2}}  \nonumber \\
&\times& \int_{-1}^{1}d{\rm cos}\theta \,{\frac{P_{r}(x^\prime_{2})P_{m}(x_{3})}{%
E+\Delta E\delta_{\beta\Lambda_c}-p_{3}^{2}/2\mu _{3}-q_{3}^{2}/2\nu _{3}
+ i\epsilon}}; \,\,\,\, \alpha,\beta=\Sigma_c,\Lambda_c, \nonumber \\
A_{31;SI;\alpha N(\beta)}^{ns_{2}i_{2}ms_{1}i_{1}}(q_{2},q_{1};E)
&=&h_{31;SI}^{s_{2}i_{2}s_{1}i_{1}}\sum_{r}\tau
_{3;s_{2}i_{2;\alpha\beta}}^{nr}(E-q_{2}^{2}/2\nu _{2})
{\frac{q_{1}^{2}}{2}}  \nonumber \\
&\times& \int_{-1}^{1}d{\rm cos}\theta \,{\frac{P_{r}(x^\prime_{3})P_{m}(x_{1})}{%
E+\Delta E\delta_{\beta\Lambda_c}-p_{1}^{2}/2\mu _{1}-q_{1}^{2}/2\nu _{1}
+ i\epsilon}}; \,\,\,\, \alpha,\beta=\Sigma_c,\Lambda_c, \nonumber \\
A_{13;SI;N\beta}^{ns_{1}i_{1}ms_{3}i_{3}}(q_{1},q_{3};E)
&=&h_{13;SI}^{s_{1}i_{1}s_{3}i_{3}}\sum_{r}\tau
_{1;s_{1}i_{1;NN}}^{nr}(E+\Delta E\delta_{\beta\Lambda_c}-q_{1}^{2}/2\nu _{1})
{\frac{q_{3}^{2}}{2}}  \nonumber \\
&\times& \int_{-1}^{1}d{\rm cos}\theta \,{\frac{P_{r}(x^\prime_{1})P_{m}(x_{3})}{%
E+\Delta E\delta_{\beta\Lambda_c}-p_{3}^{2}/2\mu _{3}-q_{3}^{2}/2\nu _{3}  
+ i\epsilon}}; \,\,\,\, \beta=\Sigma_c,\Lambda_c,
\label{gl7}
\end{eqnarray}
with the isospin and mass of particle 1 (the charmed baryon) being 
determined by the subindex $\beta$. $\mu_i$ and $\nu_i$ are the usual
reduced masses and the subindex $\alpha N(\beta)$ 
indicates a transition $\alpha N \to \beta N$ with a
nucleon as spectator followed by a $NN \to NN$ transition with $\beta$ as
spectator. $\tau_{2;s_2i_2;\alpha\beta}^{nr}(e)$ 
are the coefficients of the 
expansion in terms of Legendre polynomials of the charmed baryon-nucleon
$t-$matrix $t_{2;s_2i_2;\alpha\beta}(p_2,p_2^\prime;e)$ for the 
transition $\alpha N \to \beta N$, i.e.,
\begin{equation}
\tau_{i;s_ii_i;\alpha \beta}^{nr}(e)={2n+1 \over 2}\,{2r+1 \over 2}
\int_{-1}^1 dx_i \int_{-1}^1
dx^\prime_i\, P_n(x_i)t_{i;s_ii_i;\alpha \beta}(p_i,p^\prime_i;e)
P_r(x^\prime_i) \, .
\label{for112}
\end{equation}
The energy shift $\Delta E$, which is usually taken as $M_\alpha-M_\beta$,
will be chosen instead such that at the $\beta d$ threshold the momentum
of the $\alpha d$ system has the correct value,
i.e., 
\begin{equation}
\Delta E={[(m_\beta+m_d)^2-(m_\alpha+m_d)^2]
[(m_\beta+m_d)^2-(m_\alpha-m_d)^2]\over
8\mu_{\alpha d}(m_\beta+m_d)^2},
\label{gl11}
\end{equation}
where $\mu_{\alpha d}$ is the $\alpha d$ reduced mass.

The inhomogeneous term of Eq.~(\ref{for20}), 
$B_{2;SI;\beta}^{ns_2i_2}(q_2)$, is a two-component vector
\begin{equation}
B_{2;SI;\beta}^{ns_2i_2}(q_2) = \left( \matrix{
B_{2;SI;\Sigma_c\beta}^{ns_2i_2}(q_2)  \cr
B_{2;SI;\Lambda_c\beta}^{ns_2i_2}(q_2) \cr } \right),
\label{for3}
\end{equation}
where
\begin{equation}
B_{2;SI;\alpha\beta}^{ns_2i_2}(q_2)=h_{31;SI}^{s_2i_210}\phi_d(q_2)
\sum_r \tau_{2;s_2i_2;\alpha\beta}^{nr}(E_\beta^{th}-q_2^2/2\nu_2)
P_r(x_2^\prime).
\label{for4}
\end{equation}
$h_{31;SI}^{s_2i_2s_1i_1}$ with $s_1=1$ and $i_1=0$ are the spin-isospin
transition coefficients corresponding to a charmed baryon-deuteron initial state
(see Eq. (30), of Ref.~\cite{Ter06}), $\phi_d(q_2)$ is the deuteron wave
function, $E_\beta^{th}$ is the energy of the $\beta d$ threshold,
$P_r(x_2^\prime)$ is a Legendre polynomial
of order $r$, and
\begin{equation}
x_2^\prime={{\eta_2\over m_3}q_2 - b\over {\eta_2\over m_3}q_2 + b}.
\label{for5}
\end{equation}

Finally, after solving the inhomogeneous set of equations~(\ref{for20}),
the $\beta d$ scattering length is given by
\begin{equation}
A_{\beta d}=-\pi\mu_{\beta d}T_{\beta\beta},
\label{for6}
\end{equation}
with
\begin{equation}
T_{\beta\beta}=2\sum_{ns_2i_2}h_{13;SI}^{10s_2i_2}\int_0^\infty
q_2^2 dq_2 \phi_d(q_2)P_n(x_2^\prime) T_{2;SI;\beta\beta}^{ns_2i_2}(q_2).
\label{for7}
\end{equation}
In the case of the $\Sigma_c NN$ system, even for energies below the $\Sigma_c d$ threshold, 
one encounters the three-body singularities of the
$\Lambda_c NN$ system so that to solve the integral equations~(\ref{for20})
one has to use the contour rotation method where the momenta are 
rotated into the complex plane $q_i\to q_i e^{-i\phi}$ since as pointed
out in Ref.~\cite{Ter06} the results do not depend on the contour rotation
angle $\phi$. 
\begin{table}[b]
\begin{tabular}{cccccc}
\hline
$I$ & $J$ & $(i_{\Sigma_c},s_{\Sigma_c})$  & $(i_{\Lambda_c},s_{\Lambda_c})$ 
& $(i_{N(\Sigma_c)},s_{N(\Sigma_c)})$ & $(i_{N(\Lambda_c)},s_{N(\Lambda_c)})$ \\ 
\hline\hline
0 & 1/2 & (1/2,0),(1/2,1)  & (1/2,0),(1/2,1)  & (1,0) & (0,1) \\ 
1 & 1/2 & (1/2,0),(3/2,0),(1/2,1),(3/2,1)  & (1/2,0),(1/2,1) & (0,1),(1,0) &
(1,0)  \\ 
2 & 1/2 & (3/2,0),(3/2,1) & & (1,0) &  \\ 
0 & 3/2 & (1/2,1)  & (1/2,1) & & (0,1) \\ 
1 & 3/2 & (1/2,1),(3/2,1)  & (1/2,1) & (0,1) & \\ 
2 & 3/2 & (3/2,1) &  & & \\ \hline
\end{tabular}
\caption{Two-body $\Sigma_c N$ channels $(i_{\Sigma_c},s_{\Sigma_c})$,
$\Lambda_c N$ channels $(i_{\Lambda_c},s_{\Lambda_c})$,
$NN$ channels with $\Sigma_c$ spectator $(i_{N(\Sigma_c)},s_{N(\Sigma_c)})$, and
$NN$ channels with $\Lambda_c$ spectator $(i_{N(\Lambda_c)},s_{N(\Lambda_c)})$ that contribute to
a given $\Sigma_c NN - \Lambda_c NN$ state with total isospin $I$ and spin $J$.}
\label{tab0}
\end{table}

In order to solve the integral equations~(\ref{for20}) for the
coupled $\Sigma_c NN - \Lambda_c NN$ system we consider all configurations
where the baryon-baryon subsystems are in an $S-$wave and the third 
particle is also in an $S-$wave with respect to the pair. 
However, to construct the two-body $t-$matrices that serve as
input of the Faddeev equations we considered the
full interaction including the contribution
of the $D-$waves and of course the coupling between the $\Sigma_c N$ and
$\Lambda_c N$ subsystems (this is known as the truncated $t-$matrix
approximation~\cite{Ber86}). This approximation 
in the case of the $NNN$ system with the $NN$ interaction taken
as the Reid soft-core potential leads to a triton binding
energy which differs less than 1 MeV from the exact value~\cite{Har72}. 
We give in Table~\ref{tab0} the two-body channels
that are included in our calculation. For a given three-body state
$(I,J)$ the number of two-body channels that enter is determined by
the triangle selection rules $|J-{1\over 2}| \le s_i \le J+{1\over 2}$ 
and $|I-{1\over 2}| \le i_i \le I+{1\over 2}$. For the parameter
$b$ we found that $b=3$ fm$^{-1}$ leads to very stable results while 
for the expansion~(\ref{for11}) we took twelve Legendre polynomials,
i.e., $0\le n \le 11$. 

The two-body interactions are obtained from the chiral constituent quark model of Ref.~\cite{Val05}.
The $NN$ potentials perfectly describe the S-wave phase shifts~\cite{Gar99} and had also been used
in the study of the $\Lambda NN -\Sigma NN$ coupled channel problem~\cite{Gar07,Ter06}.
The charmed baryon-nucleon potential is derived as explained
in Ref.~\cite{Car12}. For the case of heavy
quarks ($c$ or $b$) chiral symmetry is explicitly broken and therefore boson exchanges do not
contribute. The absence of strange quarks also eliminates the contribution of $K$ or $\kappa$
exchanges as compared to the strange baryon case. This simplifies a lot the interaction, 
see Eqs.~$(8)-(13)$ of Ref.~\cite{Car12},
and gives rise to a potential whose only free parameter would be the harmonic oscillator
width of the charm quark. We will present our results for different values of $b_c$ to
get parameter free predictions. No bound states are found for the charmed two-body 
subsystems. 
\begin{table}[b]
\caption{$\Lambda_c d$, $A_{0,3/2}$ and $A_{0,1/2}$, and $\Sigma_c d$,
$A_{1,3/2}^\prime$ and $A_{1,1/2}^\prime$, scattering lengths, in fm.}
\label{t2}
\begin{tabular}{ccccccc} \hline
$A_{0,3/2}$ && $A_{0,1/2}$ && $A_{1,3/2}^\prime$ && $A_{1,1/2}^\prime$  \\ 
\hline\hline
 $-$10.27 &\,\,\,\,& 9.02&\,\,\,\, & 0.74+$\, i\, $0.18 &\,\,\,\,& 2.08+$\, i\, $0.47\\ \hline
\end{tabular}
\label{tab2}
\end{table}

There are several facts that should be noted before presenting our results
and must be considered for our discussion.
The kinetic energy associated to the $\Lambda_c^+$ is reduced compared with
that of the $\Lambda$, what would imply that the $\Lambda_c^+$ would be 
more strongly bound than the $\Lambda$ in the case of having identical interactions
with the nucleons. However the charmed baryon-nucleon interaction is weaker than that
of the strange sector due to the absence of the strange boson exchanges as also
noted in Ref.~\cite{Ban83}. Finally, note that he $\Lambda_c^+$ has a positive charge, 
whereas the $\Lambda$ is neutral. Therefore, Coulomb effects may play a non-negligible 
role in charmed nuclei. In fact, the hypertriton would be unbound if the $\Lambda$ were 
to have a positive charge. 

Bearing these considerations in mind, we have firstly proceeded to solve 
the Faddeev equations for the $\Lambda_c NN$ and
$\Sigma_c NN$ systems using the charmed baryon-nucleon and nucleon-nucleon
interactions derived from the chiral constituent quark model with 
full inclusion of the $\Lambda_c \leftrightarrow \Sigma_c$ conversion.
Let us first present the results for the $\Lambda_c d$ and $\Sigma_c d$ 
scattering lengths, that are shown in Table~\ref{tab2}. 
Although it might be difficult that they can be measured in the 
near future, however they neatly informed us about the possible
existence of bound states in the different channels. 
The $\Sigma_c d$ scattering lengths are 
complex since the inelastic $\Lambda_c NN$ channels are always open.
Both scattering lengths $A_{1,1/2}^\prime$ and $A_{1,3/2}^\prime$, have a 
positive real part indicating that the interaction is repulsive.
The spin $1/2$ $\Lambda_c d$ scattering length, $A_{0,1/2}$, 
is also positive. However the spin $1/2$ $\Lambda_c d$ scattering 
length, $A_{0,3/2}$, is negative, giving rise
to a bound state with an energy of 271 keV.

\begin{figure*}[t]
\resizebox{8.cm}{12.cm}{\includegraphics{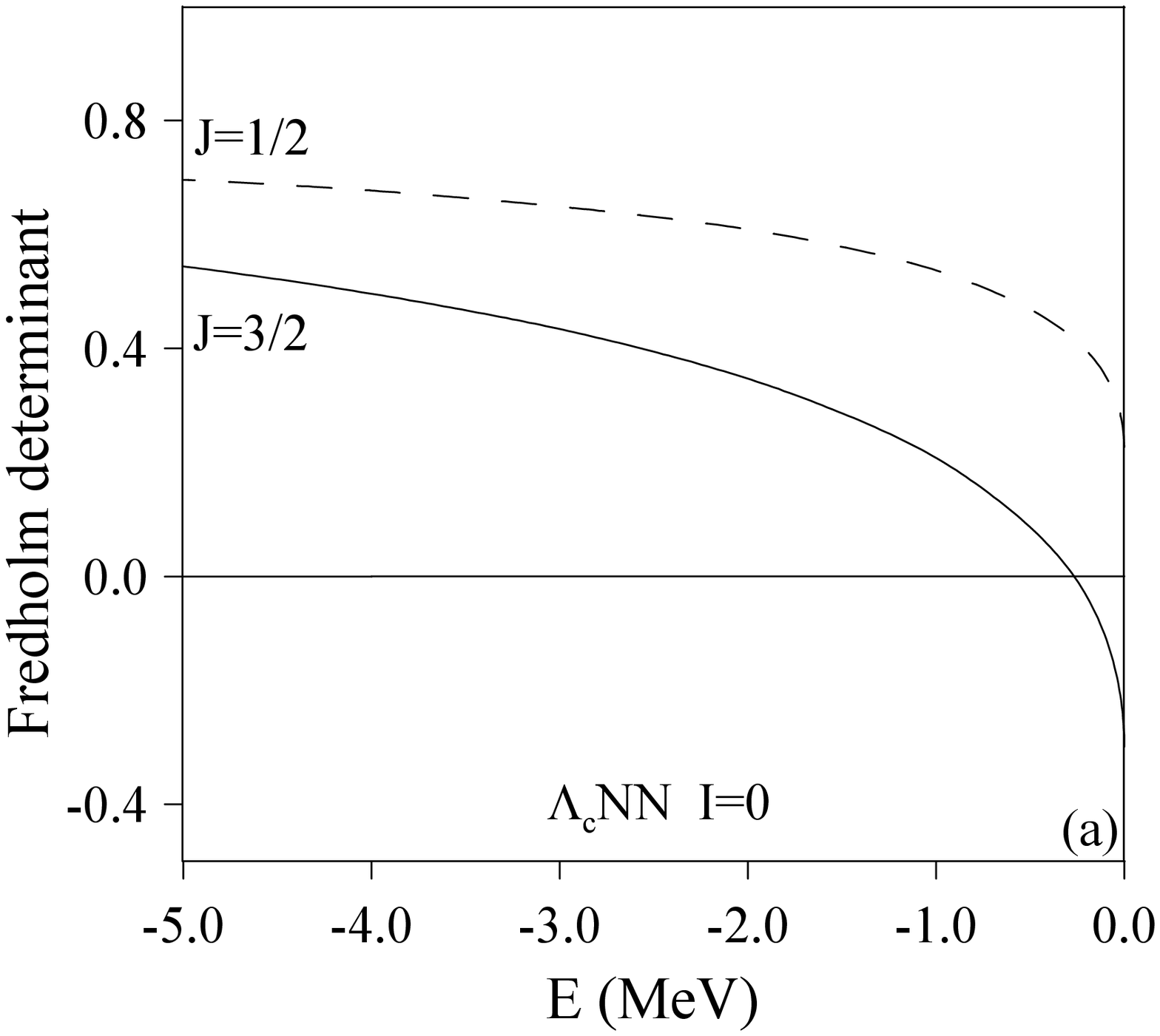}}
\resizebox{8.cm}{12.cm}{\includegraphics{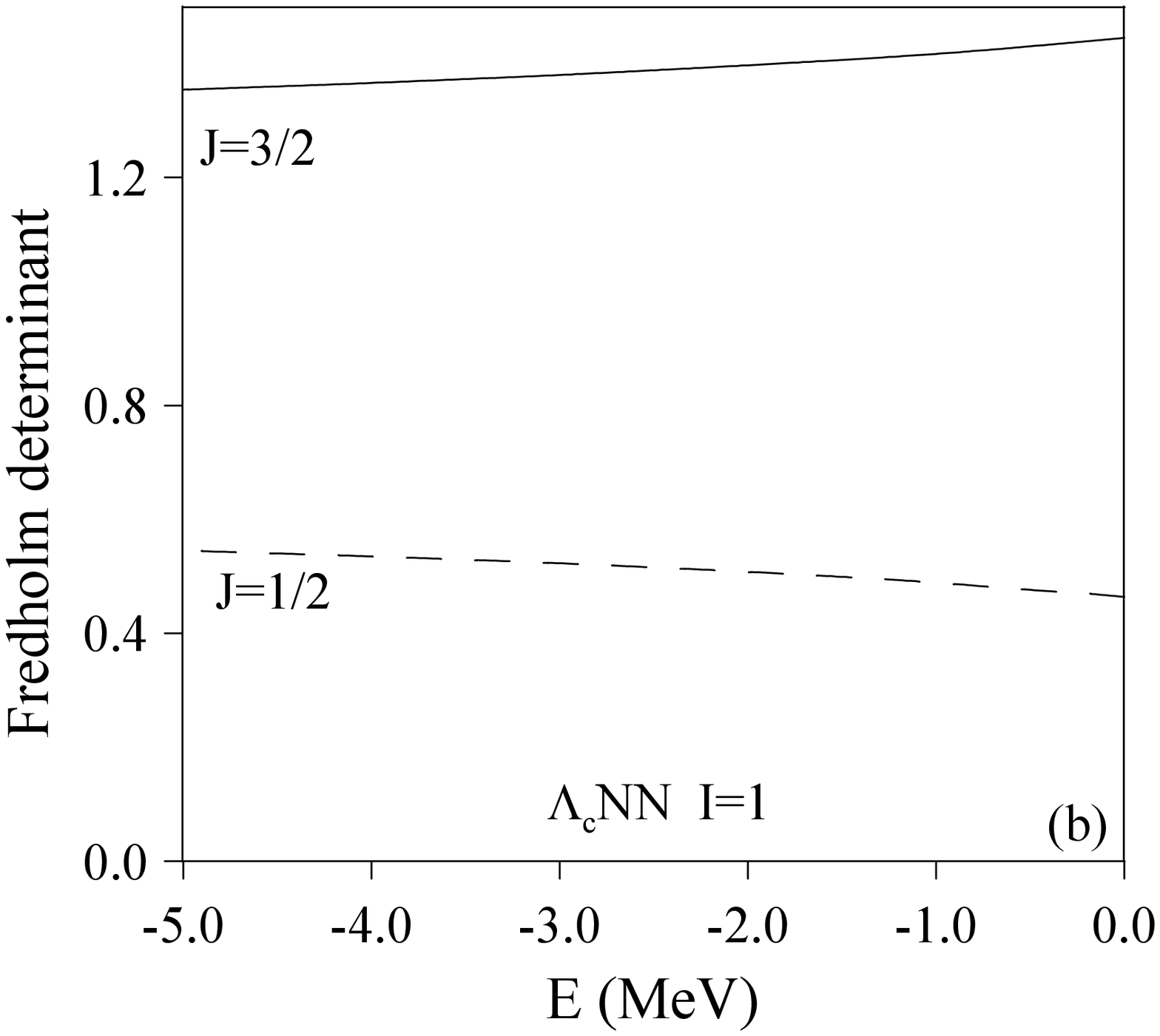}}
\vspace*{-5.0cm}
\caption{(a) Fredholm determinant for the $J=1/2$ and $J=3/2$ $I=0$
$\Lambda_c NN$ channels.
(b) Fredholm determinant for the $J=1/2$ and $J=3/2$ $I=1$
$\Lambda_c NN$ channels.}
\label{fig1}
\end{figure*}
\begin{figure*}[t]
\resizebox{8.cm}{12.cm}{\includegraphics{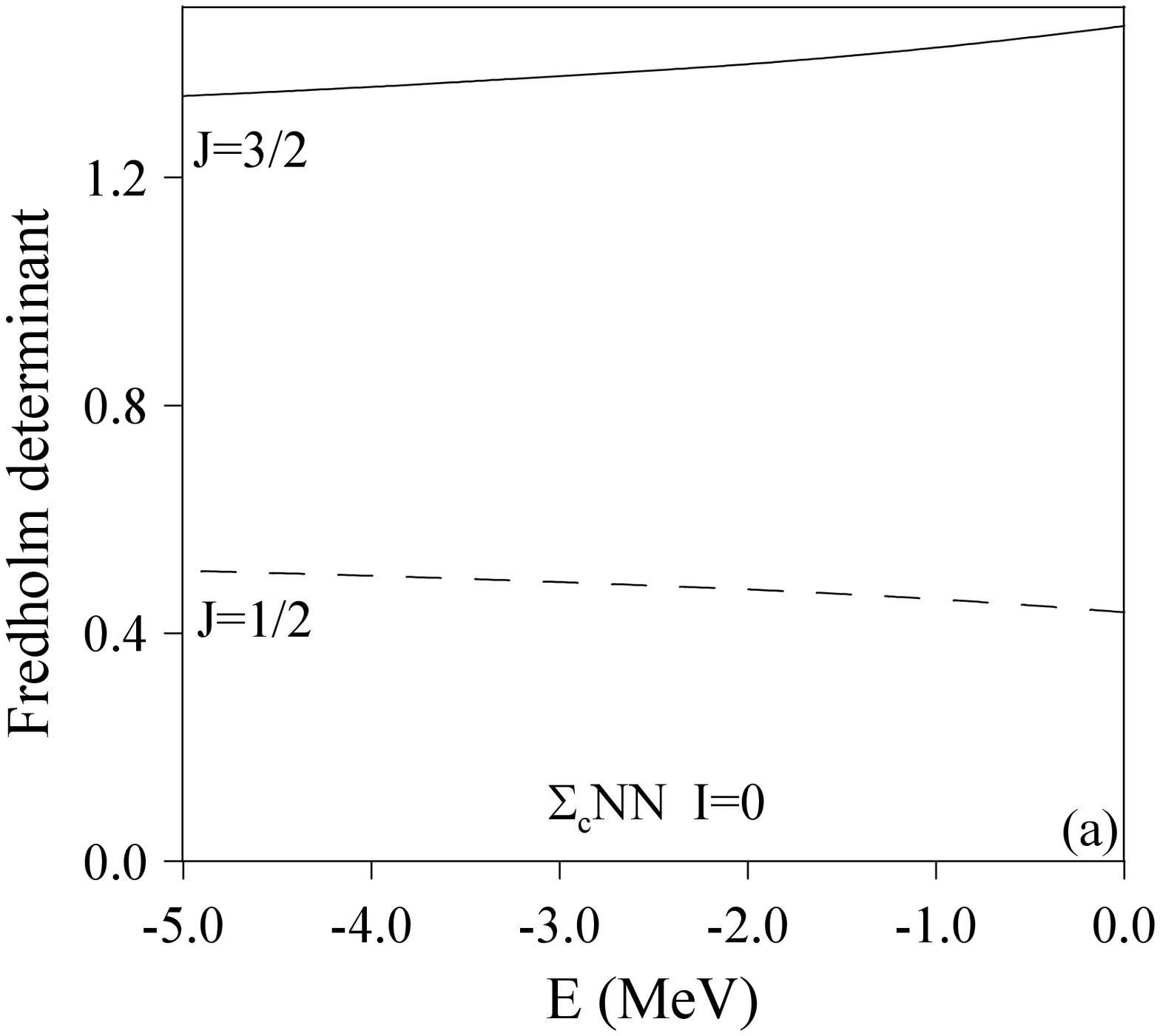}}
\resizebox{8.cm}{12.cm}{\includegraphics{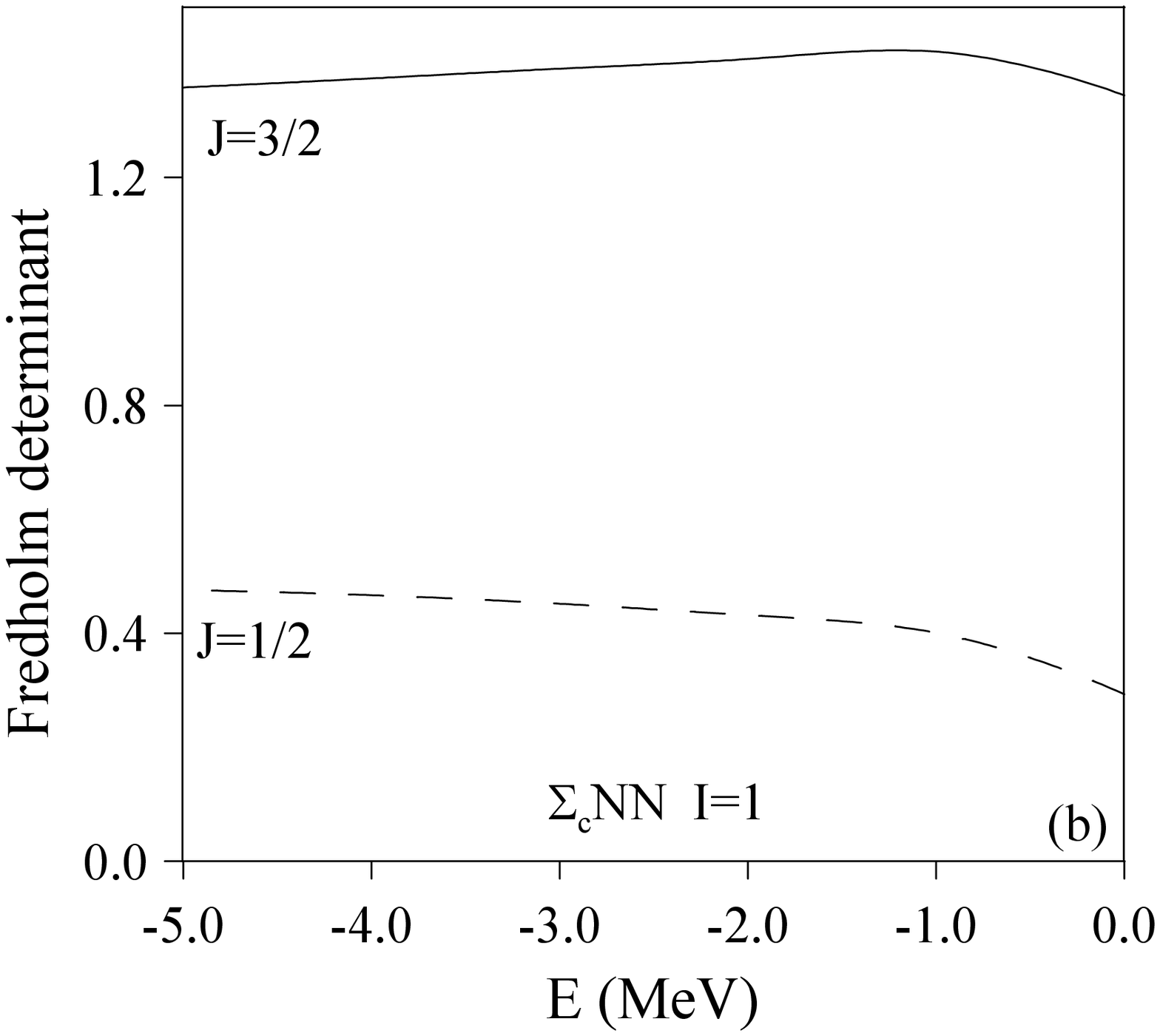}}\\\vspace*{-4cm}
\resizebox{8.cm}{12.cm}{\includegraphics{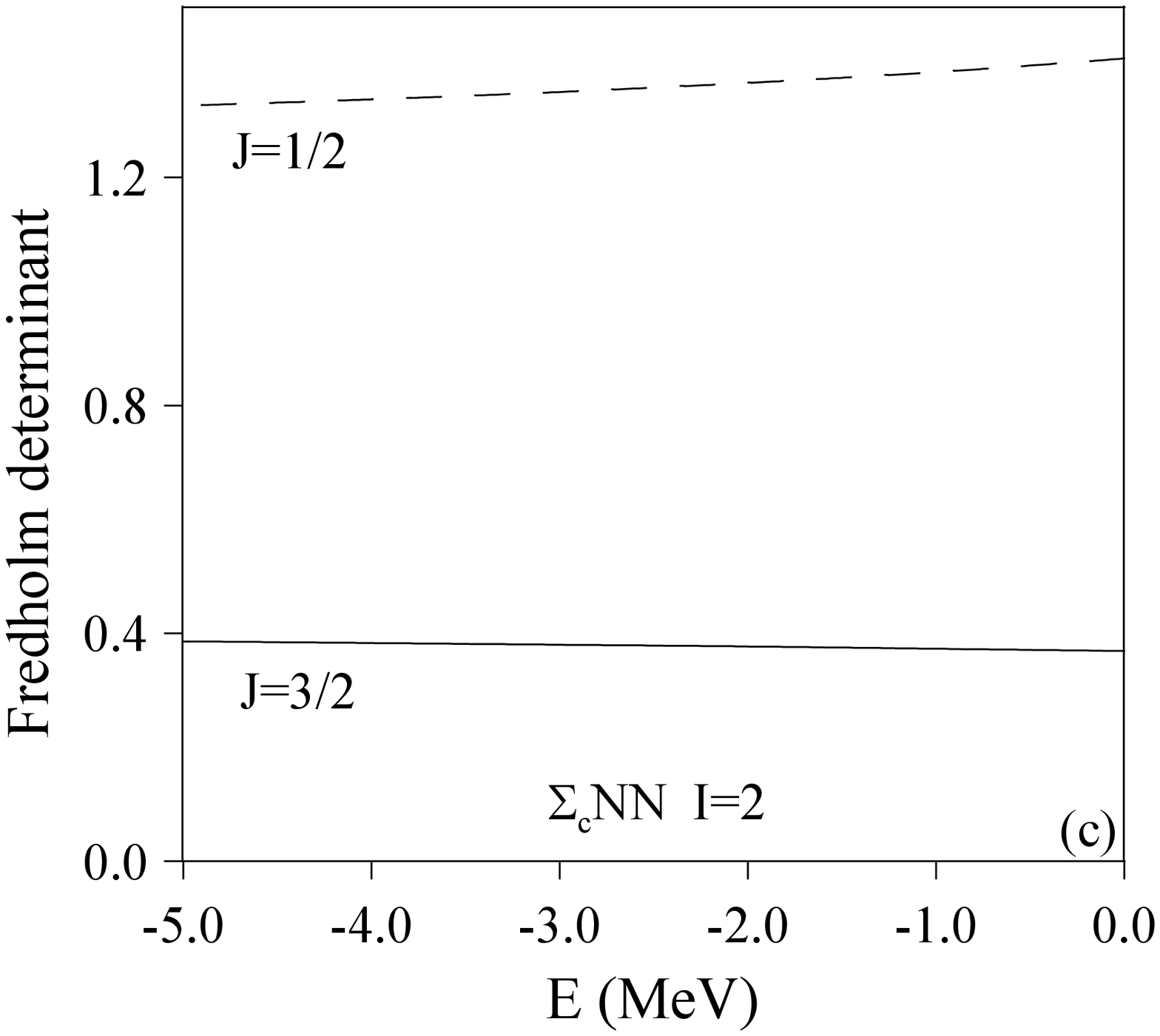}}
\vspace*{-5.0cm}
\caption{(a) Fredholm determinant for the $J=1/2$ and $J=3/2$ $I=0$
$\Sigma_c NN$ channels.
(b) Fredholm determinant for the $J=1/2$ and $J=3/2$ $I=1$
$\Sigma_c NN$ channels.
(c) Fredholm determinant for the $J=1/2$ and $J=3/2$ $I=2$
$\Sigma_c NN$ channels.}
\label{fig2}
\end{figure*}
\begin{figure*}[t]
\vspace*{-0.25cm}
\resizebox{9.cm}{13.cm}{\includegraphics{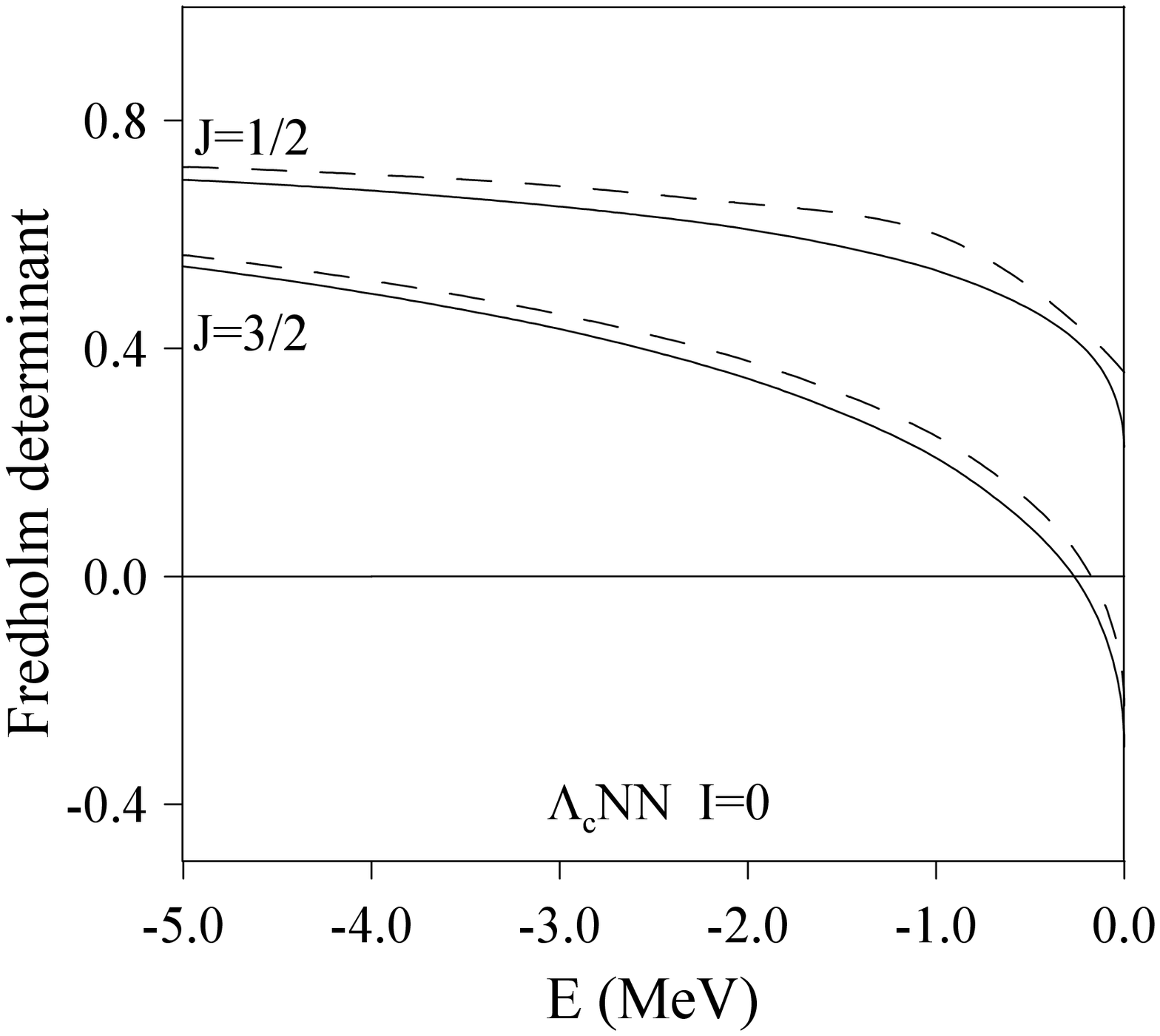}}
\vspace*{-5.5cm}
\caption{Fredholm determinant for the $J=1/2$ and $J=3/2$ $I=0$
$\Lambda_c NN$ channels. The solid line stands for the full
calculation and the dashed line for when the $\Lambda_c \leftrightarrow \Sigma_c$
transition is taken to be zero.}
\label{fig3}
\end{figure*}

We show in Fig.~\ref{fig1} the Fredholm
determinant of the different $\Lambda_c NN$ $(I,J)$ 
states~\footnote{In all $I=0$ and $I=1$ cases the zero energy corresponds to the binding energy of the deuteron
below the corresponding threshold, $\Lambda_c NN$ or $\Sigma_c NN$. The deuteron is
perfectly reproduced by our model for the $NN$ interaction~\cite{Gar99}. For $I=2$ the zero energy corresponds
to the $\Sigma_c NN$ mass because the deuteron channel does not contribute (see Table~\ref{tab0}).}.
The $I=1$ channels are repulsive, only the $I=0$ channels present attraction.
Curiously, as already noted in the scattering lengths, the order of $I=0$ channels is reversed with respect to the strange
sector, being the $J=3/2$ the most attractive one. This difference
can be easily understood due to the importance of the $\Lambda \leftrightarrow \Sigma$
conversion in the strange sector~\cite{Miy95}. 
The contribution of the $\Lambda \leftrightarrow \Sigma$ conversion
should be even stronger than $\Delta \leftrightarrow N$ conversion
in ordinary nuclei, because it is not suppressed in $S-$waves and
the $\Lambda - \Sigma$ mass difference is much smaller.
When the $N \Lambda \leftrightarrow N \Sigma$ potential
is disconnected, the $J=3/2$ channel is almost not modified while the $J=1/2$
losses great part of its attraction. Thus, the ordering between the $J=1/2$ and $J=3/2$ 
channels is reversed in such a way that the hypertriton would
not be bound (see Fig. 6(a) of Ref.~\cite{Gar07}). The $\Lambda_c \leftrightarrow \Sigma_c$ conversion is less
important than in the strange sector firstly due to their mass difference,
168 MeV as compared to the 73 MeV of the strange sector.
Besides, it comes reduced with respect to the strange sector due to 
the absence of the strange meson exchanges~\cite{Ban83}, giving rise to a 
smaller $N\Lambda_c \leftrightarrow N\Sigma_c$ transition potential. 
One should also have in mind that in the $(I,J)=(0,3/2)$ channel 
the charmed baryon-nucleon interaction with spin-singlet does not contribute, being much 
more repulsive than the spin-triplet one. Note that the ratio for the relative contribution
of the spin-singlet to the spin-triplet partial waves comes determined from the
strange sector through the $\Lambda p$ scattering cross section 
and the hypertriton binding energy~\cite{Gar07}.

We show in Fig.~\ref{fig2} the real part of the Fredholm determinant of the six
$(I,J)$ $\Sigma_c NN$ channels that are possible for energies below the 
$\Sigma_c d$ threshold. The imaginary part of the Fredholm determinant
is small. As one can see all channels are repulsive
and thus uninteresting from the point of view of possible bound states.
The larger attraction is found in the $(I,J)=(1,1/2)$ $\Sigma_c NN$ channel
that in the strange sector presented a quasibound state close to the three-body
threshold~\cite{Gar07}. Such $\Sigma NN$ quasibound state has been recently suggested
in $^3\rm{He}(K^-,\pi^\mp)$ reactions at 600 MeV/c~\cite{Har14}.

To check the relevance of the $\Lambda_c \leftrightarrow \Sigma_c$ conversion
we have solved the most interesting $\Lambda_c NN$ channels, $(I,J)=(0,1/2)$ and $(0,3/2)$,
switching off the transition between the $\Lambda_c N$ and $\Sigma_c N$ subsystems.
We plot in Fig.~\ref{fig3} the Fredholm determinant for both cases.
The solid line indicates the result of the full calculation while the
dashed one represents the results without $\Lambda_c \leftrightarrow \Sigma_c$
conversion. As can be seen the effect of the $\Sigma_c NN$ channel for $\Lambda_c NN$
is not so important as in the strange case. In particular, the order between the two channels
is not reversed. As mentioned above, the reduction of the kinetic energy associated to the 
$\Lambda_c^+$ compared with that of the $\Lambda$, implies that the $(I,J)=(0,3/2)$
state is more strongly bound than in the strange sector. 
\begin{figure*}[b]
\vspace*{-0.25cm}
\resizebox{9.cm}{13.cm}{\includegraphics{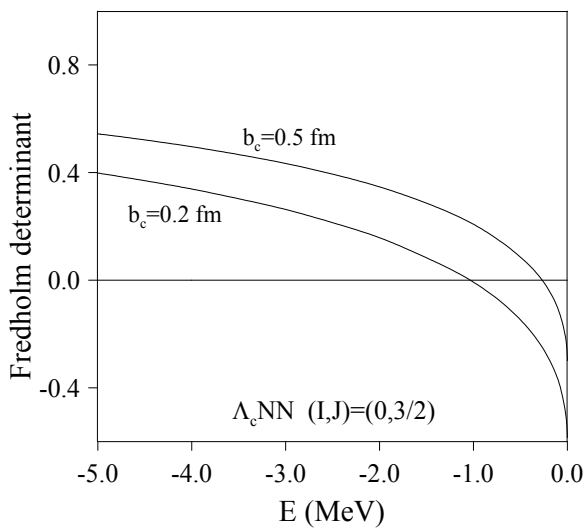}}
\vspace*{-5.5cm}
\caption{Fredholm determinant for the $(I,J)=(0,3/2)$
$\Lambda_c NN$ channel for the different values of the width
parameter for the charmed quark wave function.}
\label{fig4}
\end{figure*}

Two final remarks are in order. First, there are estimations on the literature 
about the binding energy of heavier charmed hypernuclei~\cite{Ban83} and its 
dependence on the hard core of the $N\Lambda_c$ interaction. Our quark-model 
approach presents the advantage of having the hard-core radius fixed by means 
of quark-antisymmetrization. Second, we have calculated the contribution of the
Coulomb potential exactly obtaining a contribution of $E_C=$ 131 keV, what 
would give a final binding energy of 140 keV for the $J=3/2$ charmed
hypertriton. This result could be easily understood by considering the 
$\Lambda np$ state as a bound state of a deuteron and a $\Lambda$. The 
Coulomb energy would come given at first order by,
\begin{equation}
E_C=\frac{\left\langle \Psi(\vec{r}\,)\right| V(r) \left| \Psi(\vec{r}\,) \right\rangle}
{\left\langle \Psi(\vec{r}\,)\right.  \left| \Psi(\vec{r}\,) \right\rangle} \, ,
\end{equation}
with $V(r)=\frac{\alpha}{r}$. For small binding energies $B$, the wave function can
be represented by $\Psi(\vec{r}\,)=e^{-k r}$, with $k=\sqrt{2 \eta B}$, where $\eta$ is the reduced
mass of the $\Lambda - d$ system and $B$ is the binding energy. This would give rise to
$E_C=\alpha \, k$, and using the binding energy $B$, we would obtain $E_C=$ 172 keV, comparable to
our exact result and that would not be enough as to destroy the bound state. 

To check the dependence of the binding energy of the $J=3/2$ charmed hypertriton on the
free parameter of the quark-model charmed baryon-nucleon interaction, we show in 
Fig.~\ref{fig4} the results of the $\Lambda_c NN$
$(I,J)=(0,3/2)$ state for several models with slightly different width
parameter for the charmed quark wave function. One should have in mind that 
in Ref.~\cite{Car11} it was argued that the smaller values of $b_c$ are preferred 
to get consistency with calculations based on infinite expansions, as hyperspherical harmonic 
expansions~\cite{Vij09}, where the quark wave function is not postulated. 
This also agrees with simple harmonic oscillator relations $b_c=b_n\sqrt{\frac{m_n}{m_c}}$.
As can be seen the binding energy varies between $1037$ and $271$ keV, and the 
repulsive Coulomb contribution would 
vary between $322$ and $131$ keV, what makes the competition between electromagnetic and
strong contributions crucial for the existence of this state, that would have a binding energy
between 715 and 140 keV.

In summary, we have solved the Faddeev equations for the $\Lambda_c NN$ and
$\Sigma_c NN$ systems using the charmed baryon-nucleon and nucleon-nucleon
interactions derived from a chiral constituent quark model with 
full inclusion of the $\Lambda_c \leftrightarrow \Sigma_c$ conversion.
We present results for the binding energy and the
$\Lambda_c d$ and $\Sigma_c d$ scattering lengths.
As compared to the strange sector, the kinetic energy is reduced 
but the interactions are weaker. The smaller contribution of the
$\Lambda_c \leftrightarrow \Sigma_c$ conversion due to the larger mass
difference and the smaller transition potential reverses the order of the
two only attractive channels, $(I,J)=(0,1/2)$ and $(0,3/2)$, the spin--3/2
state becoming the most attractive one. After correcting for Coulomb effects 
the charmed hypertriton would have a binding energy of at least 140 keV.
The actual experimental facilities are capable of carrying experiments seeking
for theses states what would help us in our progress in the knowledge
of the baryon-baryon interaction on the heavy-flavor sector.

\acknowledgments 
This work has been partially funded by COFAA-IPN (M\'exico), 
by Ministerio de Educaci\'on y Ciencia and EU FEDER under 
Contract No. FPA2013-47443-C2-2-P and by the
Spanish Consolider-Ingenio 2010 Program CPAN (CSD2007-00042).

\end{document}